\documentclass[aps,prl,superscriptaddress,twocolumn,%
groupedaddress,nofootinbib,nobalancelastpage,nobibnotes]{revtex4}

\usepackage{amsmath}
\usepackage{amssymb}
\usepackage{graphicx}

\newcommand{\bi}{\begin{itemize}}
\newcommand{\ei}{\end{itemize}}

\newcommand{\be}{\begin{equation}}
\newcommand{\ee}{\end{equation}}
\newcommand{\bea}{\begin{eqnarray}}
\newcommand{\eea}{\end{eqnarray}}

\newcommand{\ldm}{\Delta m_{31}^2}
\newcommand{\sdm}{\Delta m_{21}^2}

\newcommand{\fig}{Fig.}

\newcommand{\Tab}{Tab.}

\newcommand{\JHFSK}{{\sf JPARC-SK}}
\newcommand{\NUMI }{{\sf NuMI}}

\newcommand{\JHFHK}{{\sf JPARC-HK}}

\newcommand{\NuFactII}{{\sf NuFact-II}}
\newcommand{\minos}{\mbox{\sf MINOS}}
\newcommand{\icarus}{\mbox{\sf ICARUS}}
\newcommand{\opera}{\mbox{\sf OPERA}}

\newcommand{\figu}[1]{\fig~\ref{fig:#1}}

\begin{document}

\title{Is there maximal mixing in the lepton sector?}

\author{Stefan Antusch}
\affiliation{Department of Physics and Astronomy, 
University of Southampton, Southampton, SO17 1BJ, United Kingdom} 

\author{Patrick Huber}
\affiliation{Physik-Department T30d,
Technische Universit{\"a}t M{\"u}nchen, James-Franck-Stra\ss{}e,
85748 Garching, Germany}

\author{J{\"o}rn Kersten}
\email[E-mail address: ]{jkersten@ph.tum.de}
\affiliation{Physik-Department T30d,
Technische Universit{\"a}t M{\"u}nchen, James-Franck-Stra\ss{}e,
85748 Garching, Germany}

\author{Thomas Schwetz}
\affiliation{Physik-Department T30d,
Technische Universit{\"a}t M{\"u}nchen, James-Franck-Stra\ss{}e,
85748 Garching, Germany}

\author{Walter Winter}
\affiliation{Physik-Department T30d,
Technische Universit{\"a}t M{\"u}nchen, James-Franck-Stra\ss{}e,
85748 Garching, Germany}

\begin{abstract}
We discuss the potential of long-baseline neutrino
oscillation experiments to determine deviations from maximal
$\nu_\mu$-$\nu_\tau$ mixing.  We compare the obtainable sensitivities
to predictions from neutrino mass models and to the size of quantum
corrections.  We find that the theoretical expectations for deviations
are typically well within experimental reach.
\end{abstract}


\maketitle

One of the most interesting results in recent particle physics is the
evidence for large generation mixing in the lepton sector, which has
been established by neutrino oscillation experiments. Two of the three
mixing angles $\theta_{12},\theta_{23},\theta_{13}$ commonly used to
parameterize the lepton mixing matrix are large: the ``solar'' mixing
angle $\theta_{12}$ is approximately $33^\circ$, where maximal mixing
is excluded at more than $5\sigma$~\cite{Maltoni:2003da}. The best-fit
value of the ``atmospheric'' mixing angle $\theta_{23}$ is very close
to maximal mixing~\cite{Fukuda:1998mi}, where current data are still
consistent with rather large deviations from maximality: at $3\sigma$
the allowed range is $0.31 \le \sin^2\theta_{23} \le
0.72$~\cite{Maltoni:2003da}.  These results are in sharp contrast to
the quark sector, where generation mixing is small.
Maximal mixing is very interesting from the theoretical point of view,
since it corresponds to a very particular flavour structure indicating
an underlying symmetry.  On the other hand, if significant deviations
from maximality were established, the value of $\theta_{23}$ could
just be a numerical coincidence.

\begin{figure*}
\centering
\includegraphics[width=16cm]{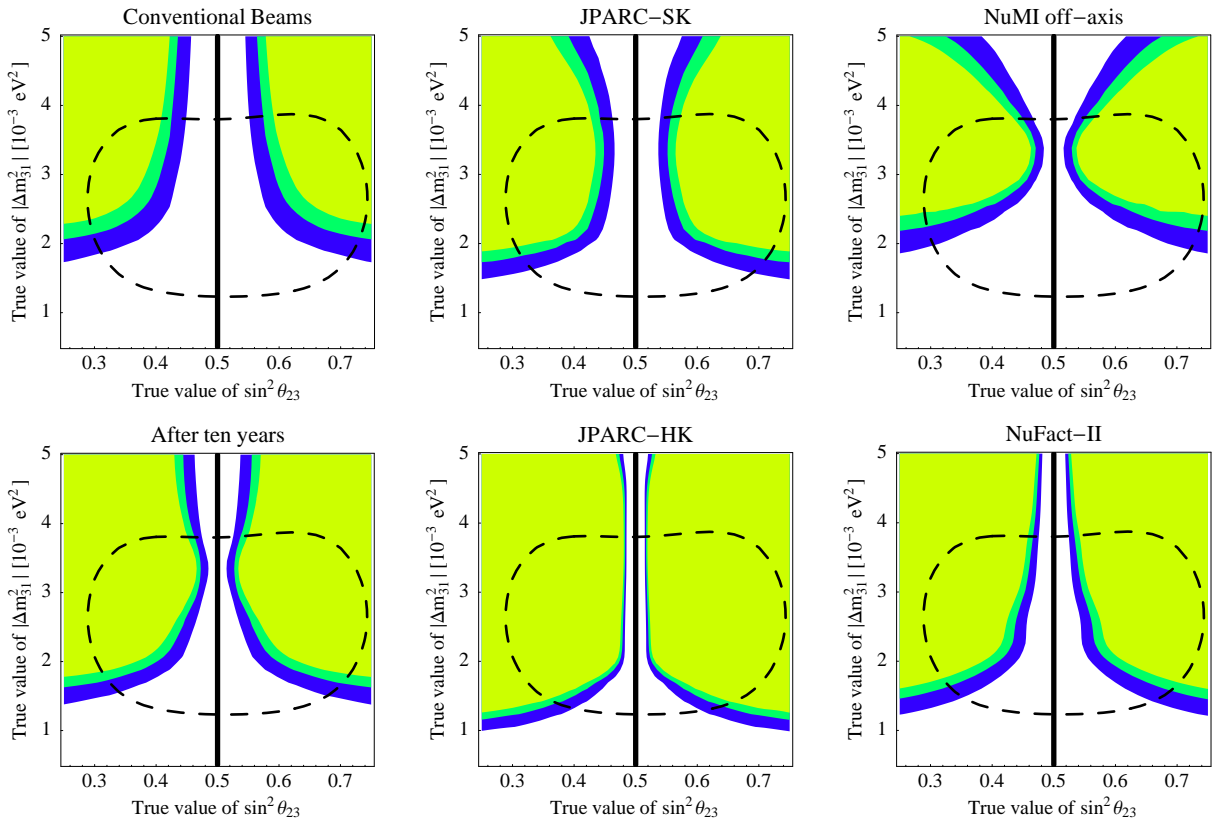}
\caption{\label{fig:summary} The regions of the true values of $\sin^2
\theta_{23}$ and $| \ldm |$ where maximal mixing can be rejected by
the considered experiment(s) at $1 \sigma$, $2 \sigma$ and $3 \sigma$
(from dark to light shading). The currently allowed region is shown
as the dashed curve at the $3 \sigma$ CL~\cite{Maltoni:2003da}.
The ``conventional beams'' refer to the combined \minos, \icarus, and
\opera\ experiments after five years of running time
each~\cite{Huber:2004ug}. The \JHFSK\ experimental parameters are used
as given in the LOI~\cite{Itow:2001ee} with five years of neutrino
running, and for the \JHFHK\ upgrade a target power of $4 \,
\mathrm{MW}$, a fiducial detector mass of $1 \, \mathrm{Mt}$, and two
years of neutrino running followed by six years of antineutrino
running are assumed~\cite{Huber:2002mx,Huber:2002rs}. For the \NUMI\
superbeam, we use a target power of $0.43 \, \mathrm{MW}$, a detector
mass of $50 \, \mathrm{kt}$, and five years of neutrino running at a
baseline of $812 \, \mathrm{km}$ and an off-axis angle of
$0.72^\circ$~\cite{Huber:2002rs,Huber:2004ug}. The label ``after ten
years'' refers to the combined potential of \minos, \icarus, \opera,
\JHFSK, and \NUMI~\cite{Huber:2004ug}. For the neutrino factory
\NuFactII\ we assume $5.3 \cdot 10^{20}$ useful muon decays per year,
a detector mass of $50 \, \mathrm{kt}$, a baseline of $3 \, 000 \,
\mathrm{km}$, and operation of four years with a neutrino beam and
four years with an antineutrino beam~\cite{Huber:2002mx}.
For the oscillation parameters not displayed, we choose $\theta_{13} =
0$, $\sdm=7 \cdot 10^{-5} \, \mathrm{eV}^2$, $\sin^2 2 \theta_{12} =
0.8$, and a normal mass hierarchy. In addition, we assume external
precisions of $10\%$ on each of $\sdm$ and $\sin2\theta_{12}$, as
expected from the KamLAND experiment, as well as a matter density
uncertainty of $5\%$ on the average matter
density~\cite{Ohlsson:2003ip}.}
\end{figure*}

A precise measurement for the leading atmospheric neutrino
oscillation parameters will be mainly obtained from the $\nu_\mu$
survival probability determined by future long-baseline experiments.
In addition to this disappearance channel, we include all
appearance channels available for a given experiment in the analysis,
which in some cases slightly increases the sensitivity to $\theta_{23}$.
For quantitative evaluations, we discuss the next generation of
conventional beam experiments, \minos~\cite{Ables:1995wq},
\icarus~\cite{Aprili:2002wx}, and \opera~\cite{Duchesneau:2002yq}. We
show their combined results after five years of running time each. In
addition, we investigate the potential of the first-generation
superbeams \JHFSK~\cite{Itow:2001ee} and \NUMI\
off-axis~\cite{Ayres:2002nm}. To estimate the potential after ten
years from now, we combine the conventional beams and first-generation
superbeams~\cite{Huber:2004ug}. Eventually, we consider also the \JHFHK\
superbeam upgrade~\cite{Itow:2001ee} and a representative setup for a
neutrino factory (labeled \NuFactII). The analysis techniques and
precise definitions for the discussed experiments can be found in
\cite{Huber:2002mx,Huber:2002rs,Huber:2004ug}. The most important
parameter values are also given in the caption of \figu{summary}.

In this figure, we show the potential of the discussed experiments to
exclude maximal $\theta_{23}$. We simulate data for fixed ``true
values'' of $\theta_{23}$ and $\ldm$ and test if they can be fitted by
$\theta_{23} = \pi/4$. For a fixed set $(\theta_{23}, \ldm)$ realized
by nature, one can exclude maximal mixing at a certain confidence
level (CL) if these values are within the corresponding shaded
region. Thus, one can easily read off how far $\theta_{23}$ has to be
from $\pi/4$ in order to be distinguished from it. For the current
best-fit value of $\ldm$, these results are summarized in
\Tab~\ref{tab:summary}.
From \figu{summary} and \Tab~\ref{tab:summary}, one can read off that
the best sensitivity to maximal mixing is obtained by \JHFHK:
deviations as small as $4\%$ of $\sin^2 \theta_{23}$ from maximal
mixing could be established at 90\% CL. The neutrino factory is not as
good as one may expect, since it measures far away from the
oscillation maximum. In fact, one can show that the sensitivity can be
improved by about a factor of two for baselines much longer than $3 \,
000 \, \mathrm{km}$. For all experiments the sensitivity strongly
decreases for low values of $\ldm \lesssim 2\cdot
10^{-3}\,\mathrm{eV}^2$, which is well within the current $3\sigma$
range.  In particular, because of the sharp energy spectrum the \NUMI\
superbeam could provide excellent results only in a rather narrow
region of $\ldm$ around $3\cdot 10^{-3}\,\mathrm{eV}^2$. Eventually,
if $\ldm$ is not too low, the combination of conventional beams,
\JHFSK, and \NUMI\ will provide a rather good measurement at the
10\%-level after about ten years from now.

\begin{table}[!b]
\centering
\begin{tabular}{lrr@{\hspace*{4mm}}rr}
\hline
Experiment(s) & \multicolumn{4}{c}{$|0.5 - \sin^2\theta_{23}|$} \\
& \multicolumn{2}{c}{90\% CL} & \multicolumn{2}{c}{$3 \sigma$} \\ 
\hline
Conventional beams & $0.100$ & $20\%$ & $0.148$ & $30\%$\\
\JHFSK\            & $0.057$ & $11\%$ & $0.078$ &$16\%$ \\
\NUMI\             & $0.079$ & $16\%$ & $0.126$ &$25\%$ \\
After ten years    & $0.050$ & $10\%$& $0.069$ & $14\%$\\
\JHFHK\            & $0.020$ & $4\%$& $0.024$ & $5\%$\\
\NuFactII          & $0.055$ & $11\%$& $0.075$ & $15\%$\\
\hline
\end{tabular}
\caption{\label{tab:summary} Minimal values of $|0.5 -
\sin^2\theta_{23}|$ required to exclude maximal mixing at 90\% CL and
$3 \sigma$ (absolute and relative values).  For the oscillation parameters, we use the same values as in \figu{summary} and $\ldm=2.5 \cdot 10^{-3} \,
\mathrm{eV^2}$.}
\end{table}

The results in \figu{summary} are calculated for the true value
$\theta_{13}=0$. For $\theta_{13}$ close to the current bound, none of
the shown results changes drastically. Only the neutrino factory
potential is slightly improved, since the $\nu_e \to \nu_\mu$ channel
contributes somewhat to the measurement of $\theta_{23}$.  For all
experiments, the results are basically independent of the mass
hierarchy. Note that although the sensitivity to
$|0.5-\sin^2\theta_{23}|$ is rather good, in general it is very
difficult to determine the sign
(``$\theta_{23}$-degeneracy''~\cite{Fogli:1996pv,Barger:2001yr}).
Finally, we remark that irrespective of the true value of
$\sin^2\theta_{23}$, the achievable accuracy is very similar to the
sensitivities shown in \figu{summary} and
\Tab~\ref{tab:summary}.

Let us now analyze theoretical expectations
for the deviation from maximal atmospheric mixing. It could either be a
feature of a mass model itself, or it could stem from quantum corrections
due to the running of $\theta_{23}$ between high energy, where the model is defined,
and low energy, where the experiments are performed. 
As to the first possibility, there exists a large variety of models
aiming to explain the observed neutrino properties, utilizing various
approaches such as Grand Unification, flavour symmetries, sequential
right-handed neutrino dominance, textures, or combinations of these.
Many of them are based on a version of the see-saw mechanism.  
\begin{table}
\begin{tabular}{lcr} 
\hline
Model(s) & Refs.\ & $|0.5-\sin^2\theta_{23}|$ \\
\hline
Minimal SO(10) & \cite{Goh:2003hf} & $>0.16$\\
SO(10) + flavour symmetry
 & \cite{Pati:2002pe,Blazek:1999hz,Albright:2001uh}
 & $\lesssim 0.05$ \\
SO(10) + texture & \cite{Bando:2003ei} & $\lesssim 0.11$ \\
\hline
Flavour symmetries
 & \cite{Grimus:2001ex,Grimus:2003kq,Babu:2002dz,Ohlsson:2002rb,Harrison:2003aw,Kubo:2003pd,Ma:2004yx}
 & 0 \\
\hphantom{Flavour symmetries} & \cite{Chen:2004rr} & 0.02 \\
\hphantom{Flavour symmetries} & \cite{Raidal:2004iw} & 0.04 \\
\hline
Sequential RH neutrino dominance & \cite{King:1999mb,King:2002nf}
& 0.1 \\
\hphantom{S}+ Flavour symmetries & \cite{Ross:2002fb,King:2003rf,Ross:2004qn}
 & 0.1 \\ 
\hphantom{S}+ Type II see-saw upgrade & \cite{Antusch:2004xd} & 0.01 .. 0.1 \\ 
\hline
Texture zeros & \cite{Bando:2003wb} & 0.07 \\
\hphantom{Texture zeros} & \cite{Zhou:2004wz} & $>0.1$ \\
Perturbations of textures & \cite{Frampton:2004ud} & $\lesssim 0.16$ \\
\hphantom{Perturbations of textures}
 & \cite{Rodejohann:2004cg,deGouvea:2004gr} & 0.005 .. 0.1 \\
\hline
\end{tabular}
\caption{
 Selection of theoretical expectations for $|0.5-\sin^2\theta_{23}|$ at
 tree level.
 The numbers should be considered as order of magnitude statements.
}
\label{tab:ModelPredictions}
\end{table}
There are models where the predicted $\theta_{23}$ lies in a range 
that does not include maximal mixing at all
\cite{Ross:2002fb,Goh:2003hf,Zhou:2004wz}. 
In many other cases a large atmospheric angle can be explained, while almost
maximal mixing would require some tuning, see e.g.\
\cite{Buchmuller:2001dc,Maekawa:2001uk,Appelquist:2002me,%
Bando:2003ei,Casas:2003kh,Mohapatra:2003tw,Shafi:2004jy,Dorsner:2004qb}.
Other works, for instance
\cite{Blazek:1999hz,Albright:2001uh,Pati:2002pe,Asaka:2003iy},
predict a value of $\theta_{23}$ 
rather close to $\pi/4$ at leading order, but various sources cause
deviations that are typically still within the reach of future experiments.

In many cases, these deviations are related to small parameters, such
as mass ratios.  For example, even if we assume that maximal
$\theta_{23}$ is predicted from properties of the neutrino mass
matrix, corrections can stem from the charged lepton sector, with a
typical order of magnitude of $|0.5-\sin^2\theta_{23}| = {\cal O}
(m_\mu/m_\tau) \sim 0.06$.  Analogously, assuming that maximal
$\theta_{23}$ is predicted from the charged lepton mass matrix, a
hierarchical neutrino mass matrix might induce
$|0.5-\sin^2\theta_{23}| = {\cal O} (m_2/m_3) \sim 0.17$
\cite{Antusch:2004re}.
Deviations of this order of magnitude are also typical in models based on 
sequential right-handed neutrino dominance,  
where maximal $\theta_{23}$ in leading order can originate from the dominant
right-handed neutrino and the subdominant contribution 
leads to corrections 
(see e.g.\ \cite{King:1999mb,King:2002nf,Ross:2002fb,King:2003rf,Ross:2004qn}).

Thus, the described classes of models, summarized in
\Tab~\ref{tab:ModelPredictions}, favor deviations from maximal
atmospheric mixing that will be measurable unless $\ldm$ is very
small.
If they are not found experimentally, some new ingredients will be
necessary. A value of $\theta_{23}$ very close to $\pi/4$ corresponds
to a rather particular configuration of lepton mixing parameters,
which is clearly not compatible with the assumption of a neutrino mass
matrix without any structure~\cite{Haba:2000be} and would require some
theoretical reason. One option is employing flavour symmetries that
enforce virtually maximal atmospheric mixing, see e.g.\
\cite{Grimus:2001ex,Babu:2002dz,Ohlsson:2002rb,Harrison:2003aw,Grimus:2003kq,Kubo:2003pd,Ma:2004yx}.
On the other hand, if maximal mixing is excluded experimentally by a
broad margin, this will favor either a numerical coincidence without
an underlying symmetry or models which can accommodate or even predict
significant deviations.
Either way, precise measurements of $\theta_{23}$ will provide
crucial information on the flavour structure of lepton mass models.

Models employing flavour symmetries, GUT relations or textures typically
operate at a very high energy scale.  Consequently, their predictions
are modified by radiative corrections, i.e.\ the renormalization group
(RG) running to low
energy, where experiments take place.  This means that even for
a model predicting exactly maximal
atmospheric mixing, one expects to measure deviations of the order of
magnitude of the running effects \cite{Antusch:2003kp}.  
Of course, the combination of
deviations from $\pi/4$ at high energy and quantum corrections could, in
principle, produce nearly maximal mixing at low energy. However, this
possibility appears unnatural, since it requires a conspiracy between
two effects that are not related in general.

\begin{figure*}
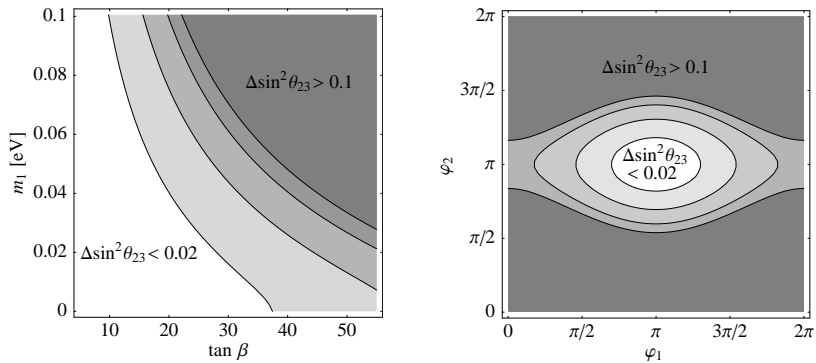

\centering
\includegraphics[scale=0.60]{RGcorrectionsT23_1}
\qquad
\includegraphics[scale=0.60]{RGcorrectionsT23_2}
\caption{
Deviations of $\sin^2\theta_{23}$ from $0.5$ due to the running in the
MSSM  between high energy $M_\mathrm{U}\approx 2 \cdot 10^{16}$ GeV, where 
maximal $\theta_{23}$ has been taken as initial condition, and low 
energy $M_\mathrm{EW} \approx 10^2$ GeV. The contour lines correspond to
deviations roughly equal to the $90\%$ CL sensitivities listed in
\Tab~\ref{tab:summary},
$\Delta \sin^2\theta_{23} = 0.02,0.05,0.08$ and $0.1$, respectively.
In the left figure, the corrections are shown as a function of $\tan \beta$ and 
$m_1$, the mass of the lightest neutrino for a normal mass scheme. The right 
plot illustrates the dependence on the Majorana CP phases $\varphi_1$ and 
$\varphi_2$ (as defined in \cite{Antusch:2003kp}) for $m_1=0.075$ eV. We have used 
$\Delta m^2_{31}=2.5 \cdot 10^{-3}$~eV$^2$
and the same values as in \figu{summary} for the other 
parameters as further boundary conditions.    
}
\label{fig:RG}
\end{figure*}

One can easily estimate the size of the RG effects using the
differential equation for $\theta_{23}$ derived in \cite{Antusch:2003kp}.
It immediately follows
that the effects are negligible in the Standard Model due to the
smallness of the charged lepton Yukawa couplings.  In the MSSM, these
are enhanced by $\tan\beta$, the ratio of the two Higgs vevs, so that
the situation can change.
In addition to the oscillation parameters,
the running depends on the mass of the lightest neutrino, the value of
the Majorana CP phases in the lepton mixing matrix, and $\tan\beta$.  
The MSSM results are shown in Fig.~\ref{fig:RG}.
For a considerable parameter range, one finds
corrections to $\theta_{23}$ comparable to the precision of future experiments.
Note that this is a conservative estimate, as we have neglected
additional contributions coming from neutrino Yukawa couplings above
the see-saw scale \cite{King:2000hk,Antusch:2002rr,Antusch:2002hy},
which can cause sizable effects even in the Standard Model.
Physics above the GUT scale could also contribute \cite{Vissani:2003aj}.
This provides a further argument why precision experiments have a good
chance of measuring deviations from maximal atmospheric mixing.

In summary, we have discussed the potential of future long-baseline
experiments to test maximal atmospheric mixing.  The
comparison with fermion mass models has shown that the deviations from 
maximal mixing predicted by many of them are large enough to be
experimentally accessible. 
We have furthermore discussed the effects of renormalization group
running, which connects the models built at very high energy scales with the measurements at low energies. These effects are also likely to cause deviations from maximality accessible by planned experiments. We conclude that if no deviation from maximal mixing can be established, the models will be severely constrained. This result will point towards a symmetry for maximal $\theta_{23}$ and indicate small quantum corrections.
Finally, compared to experiments involving quarks, measurements in the
leptonic sector do not suffer from the limitation by hadronic
uncertainties.  Thus, in the long term, the combination of precision
measure\-ments of the atmospheric angle and other neutrino parameters,
such as $\theta_{13}$, has the potential to play 
an important role for exploring GUT-scale physics.

This work was supported in part by the  
``Son\-der\-forschungs\-bereich~375 f\"ur Astro-{}Teil\-chen\-phy\-sik der 
Deut\-schen Forschungsgemeinschaft''. S.A.\ acknowledges support from the 
PPARC grant PPA/G/O/2002/00468. 
We would like to thank Steve King, Manfred Lindner, Michael Ratz and
Mark Rolinec for useful discussions.

{\footnotesize
\bibliography{./references}
}

\end{document}